\documentclass[authoryear,preprint,review,12pt]{elsarticle}

\usepackage{amssymb}
\usepackage[utf8]{inputenc}
\usepackage{standalone}
\usepackage{subfig}
\usepackage{url}
\usepackage{geometry}
\usepackage[table]{xcolor}
\usepackage{siunitx}

\journal{Cold Region Science and Technology}

\begin{document}

\begin{frontmatter}



\title{Towards the assimilation of satellite reflectance into semi-distributed ensemble snowpack simulations\footnote{\includegraphics[scale=0.25]{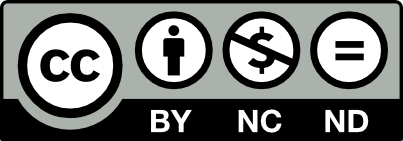} \copyright 2019 by the author(s). Distributed under a Creative Commons CC-BY-NC-ND licence.}}



\author[label1]{Bertrand Cluzet}
\ead{bertrand.cluzet@meteo.fr}
\author[label1]{Jesus Revuelto}
\author[label1]{Matthieu Lafaysse}
\author[label1,label2]{François Tuzet}
\author[label2]{Emmanuel Cosme}
\author[label2]{Ghislain Picard}
\author[label2]{Laurent Arnaud}
\author[label1]{Marie Dumont}

\address[label1]{Univ. Grenoble Alpes, Université de Toulouse, Météo-France, CNRS, Centre d'Études de la Neige, Grenoble, France}
\address[label2]{Institut des Géosciences de l’Environnement, IGE, UGA-CNRS, Grenoble, France}
\begin{abstract}
Uncertainties of snowpack models and of their meteorological forcings limit their use by avalanche hazard forecasters, or for glaciological and hydrological studies.
The spatialized simulations currently available for avalanche hazard forecasting are only assimilating sparse meteorological observations. 
As suggested by recent studies, their forecasting skills could be significantly improved by assimilating satellite data such as snow reflectances from satellites in the visible and the near-infrared spectra. Indeed, these data can help constrain the microstructural properties of surface snow and light absorbing impurities content, which in turn affect the surface energy and mass budgets. This paper investigates the prerequisites of satellite data assimilation into a detailed snowpack model. An ensemble version of Météo-France operational snowpack forecasting system (named S2M) was built for this study.
This operational system runs on topographic classes instead of grid points, so-called 'semi-distributed' approach. Each class corresponds to one of the 23 mountain massifs of the French Alps (about 1000km\textsuperscript{2} each), an altitudinal range (by step of 300m) and aspect (by step of 45\textsuperscript{o}). We assess the feasability of satellite data assimilation in such a semi-distributed geometry. Ensemble simulations are compared with satellite observations from MODIS and Sentinel-2, and with in-situ reflectance observations. The study focuses on the 2013-2014 and 2016-2017 winters in the Grandes-Rousses massif.
Substantial Pearson R\textsuperscript{2} correlations (0.75-0.90) of MODIS observations with simulations are found over the domain. This suggests that assimilating it could have an impact on the spatialized snowpack forecasting system. However, observations contain significant biases (0.1-0.2 in reflectance) which prevent their direct assimilation. MODIS spectral band ratios seem to be much less biased. This may open the way to an operational assimilation of MODIS reflectances into the Météo-France snowpack modelling system.
\end{abstract}


\end{frontmatter}


\section*{Highlights}
\label{sec:high}

- Ensemble simulations of the snowpack are compared with satellite reflectances\\
- Spatial aggregation into the semi-distributed geometry filters the observation noises\\
- Satellite reflectances carry useful information worth to assimilate\\
- MODIS reflectances can not be directly assimilated because they are biased\\
- Ratios of MODIS reflectances show no evidence of bias and could be assimilated\\

\section{Introduction}
\label{sec:intro}

The avalanche forecasting services of some countries use a chain composed of meteorological forcings, coming from either a Numerical Weather Prediction model (NWP) or observations, and a detailed multilayer snowpack model such as Crocus \citep{vionnet2012} or SNOWPACK \citep{lehning2002}. Both meteorological forcings and snowpack modelling induce errors and uncertainties in the simulations \citep{essery2013,vernay2015,raleigh2015,gunther2019uncertainties}. These errors are considerably limiting the use of snowpack models by avalanche hazard forecasters \citep{morin2018issw}. The representativeness of simulations is also limited in complex mountain terrain \citep{fiddes2012}. In addition, most of these snowpack modelling chains do not operationally assimilate any available information on the snowpack properties (either in-situ or remotely-sensed) \citep{helmert2018review}. There are several reasons for that : (1) snowpack in-situ observations are sparse and lack representativeness (2) satellite observations retrieval is challenging \citep{nolin2011,helmert2018review}, (3) preserving state variable consistency within detailed snowpack models, which is a key point for avalanche forecasting, requires sophisticated assimilation algorithms \citep{magnusson2017improving}. As a consequence, the errors often accumulate along the snow season leading to increasingly poor model performance and utility for avalanche hazard forecasting and other operational applications.\\

Data assimilation systems using ensemble approaches is the best way to reduce snowpack modelling errors \citep{charrois2016,larue2018simulation, piazzi2018particle, winstral2019bias}. The Particle Filter (PF) ensemble assimilation algorithm seems to be especially well suited to reduce detailed snowpack modelling errors \citep{magnusson2017improving}. Indeed, ensembles enable to quantify the uncertainties of (1) meteorological forcings, using physically based ensembles \citep{vernay2015} or statistical perturbations \citep{charrois2016,winstral2019bias}, and (2) snowpack modelling, using multiphysical systems \citep{essery2015,lafaysse2017multiphysical}. \cite{charrois2016} did the first application of a PF within a detailed snowpack model, but only at one specific location and their ensemble only described the meteorological uncertainty, not model uncertainty. They were followed by \cite{magnusson2017improving} and \cite{larue2018simulation}. Recently, \cite{piazzi2018particle} and \cite{smyth2019particle} applied the PF to a combination of meteorological and model ensembles, but with a less complex model and at the local scale as well. In parallel, spatialized application of PF has been done in several studies \citep{thirel2013,baba2018assimilation}, but with deterministic and low complexity snow models, not suited for avalanche hazard forecasting. This paper fills a gap by implementing a combination of a meteorological ensemble and a multiphysical system of detailed snow models in a spatialized context.\\

Daily moderate-resolution observations (250 to 500 \si{\metre}) in the visible (VIS) and near infra-red (NIR) spectrum from the MODerate Resolution Imaging Spectroradiometer (MODIS) are suitable to monitor the snowpack properties \citep{hall2002modis}. Sentinel-2 (S2) has a coarser revisit time (5 days) but captures much finer spatial scales (10-20 m). From MODIS and S2 spectral Top Of Atmosphere (TOA) radiance products, it is possible to retrieve the snowpack extent as a Snow Cover Fraction by pixel (SCF) and Bottom of Atmosphere (BOA) reflectances which requires to account for the complexity of the radiative transfer in mountainous area \citep{richter1998correction,sirguey2009b}. Many studies focus on the assimilation of SCF, showing a strong impact of assimilation in hydrological models \citep{delannoy2012,thirel2013,stigter2017assimilation, aalstad2018ensemble,baba2018assimilation}. However, SCF is expected to be of less interest for detailed snowpack modelling in alpine terrain, because the information content is limited to the snow line \citep{andreadis2006assimilating,toure2018assimilation}. Meanwhile, it is expected for the BOA reflectances to carry useful information on the temporal and spatial variability of the snowpack surface properties such as Light Absorbing Particles concentration (LAP, [\si{\kilogram.\kilogram_{snow}^{-1}}]) and snow microstructure (quantified by the Specific Surface Area, SSA, [\si{\metre^2.\kilogram^{-1}}]) \citep{dozier2009interpretation,kokhanovsky2018reflectance}. Indeed, these variables drive the shortwave (SW) radiation absorption of the snowpack, and thus carry crucial information on the snow surface energy budget \citep{skiles2018radiative,dimauro2019saharan}. Moreover, monitoring the surface snow microstructure can help detect precipitation (solid and liquid) and melting events, while frequent observations of surface LAP contents can enable to constrain LAP vertical layering within the snowpack. In line with this, \cite{charrois2016} showed that assimilating satellite reflectances could help reduce Snow Water Equivalent (SWE, [\si{\kilogram.\metre^{-2}}]) modelling uncertainties by up to 45\%.\\

The most detailed snow models are also able to compute reflectances from the snowpack properties, through the use of a detailed radiative transfer \citep{libois2015,skiles2019toward} and the explicit evolution of SSA \citep{carmagnola2013a} and LAP \citep{tuzet2017multilayer}.  Such radiative transfer models play the role of observation operators, computing observation-like variables from the model state variables. However, modelling geometries often differ from the distributed geometry of satellite retrievals \citep{mary2013}. For instance, Météo-France multilayer snowpack model Crocus is operationally applied on several topographical classes (by 300 m elevation bands, for 8 different aspects and 3 different slopes, so-called "semi-distributed" geometry) inside so-called "massif" regions of about 1000 km\textsuperscript{2} \citep{durand1999,lafaysse2013}. This semi-distributed framework, with around 200 topographical classes, was proven to be sufficient to represent the main features of snowpack variability with topography compared to fully distributed simulations at 25 to 250 m resolution \citep{fiddes2012,revuelto2018multi}. However, the feasibility of the assimilation of satellite reflectances in Crocus semi-distributed model using the PF ensemble data assimilation algorithm, still needs to be assessed.\\

The main objective of this paper is to assess the potential for semi-distributed assimilation of satellite observations of snowpack reflectances into ensemble snowpack simulations. For that purpose, we present extended comparisons of openloop simulations (e.g. without assimilation) with satellite observations from MODIS and S2 aggregated in this geometry. Sec. \ref{sec:data} presents the data and the modelling framework, while Sec. \ref{sec:met} introduces the aggregation method and defines the points of comparison from the assimilation perspective. Then Sec. \ref{sec:results} presents the comparison results, which are discussed in Sec. \ref{sec:disc}.\\

\section{Data and model}
\label{sec:data}

\subsection{Case study}
\label{sec:case_study}
\begin{figure}[h!]
	\centering
	\includegraphics[width=140mm]{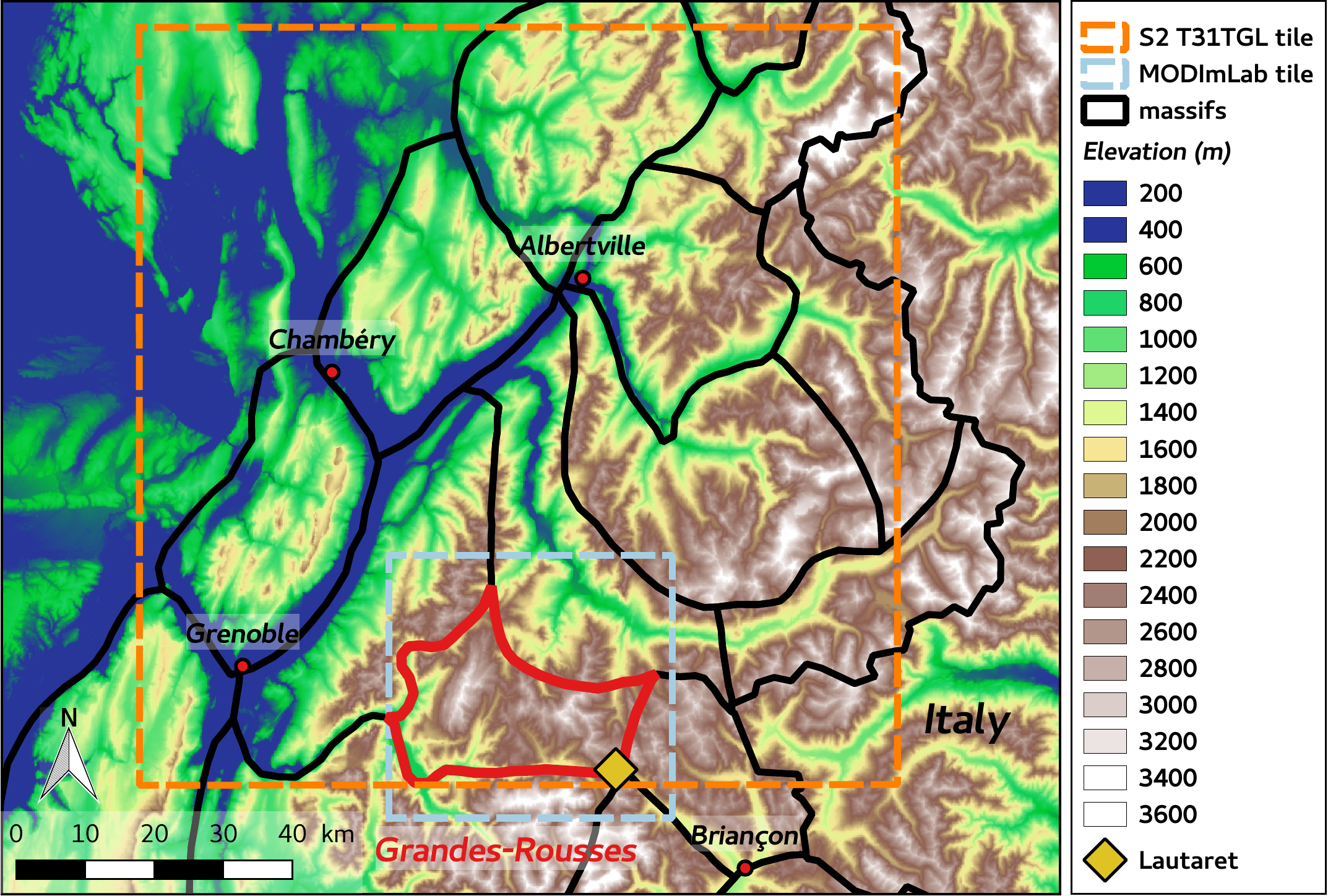}
	\caption{Map of the study area of the Grandes-Rousses (red), located in the central French Alps. Lautaret field site (diamond) and satellite retrieval tiles (boxes) are also indicated, together with the limits of other SAFRAN massifs (black). Source : Shuttle Radar Topography Mission (SRTM), resolution : 90m.}
	\label{fig:map1}
\end{figure}

This study focuses on two snow seasons (2013-2014 and 2016-2017) in the Grandes-Rousses (see Fig. \ref{fig:map1}). The area of about 500 \si{\kilo\meter^{2}} is located in the Central French Alps, and is characterized by a wide elevation range from the bottom of Romanche valley (about 700 m a.s.l.) to the top of Aiguilles d'Arve (3514 m a.s.l.). This specific massif was chosen because it encompasses the Col du Lautaret (2058 m a.s.l.), where field campaigns have been carried out since winter 2016-2017 close to an automatic weather station \citep{tuzet2019influence}.\\
The two snow seasons have been selected because they show contrasted snow conditions. 2013-2014 is characterised by above average snow depths, with frequent snowfall events and two major dust deposition events (end of February, end of March) \citep{dumont2017situ,dimauro2015mineral}. 2016-2017 was a warm winter, without significant snowfall between late November and beginning of January, and early melting in spring. In addition, several minor dust deposition events occurred after the end of February according to MOCAGE outputs.\\

\subsection{Digital Elevation Model (DEM) and landcover}

\subsubsection{DEM}
\label{sec:dem}
Digital Elevation Models (DEM) of the study area are used here to retrieve satellite data and to perform a topographical aggregation of observations into the model semi-distributed geometry. For that purpose, DEM BD Alti\textregistered \footnote{\url{http://professionnels.ign.fr/bdalti}} (IGN25) from the French Geographical Institute (\textit{Institut National de l’information Géographique et forestière}, IGN) with native 25 \si{\metre} resolution was used in this study at different scales : 125 \si{\metre} for the retrieval of MODIS images (IGN125) (see Sec. \ref{sec:modis}) and 250 \si{\metre} (IGN250) for the topographical aggregation. In addition, a different DEM from Shuttle Radar Topography Mission (SRTM, \cite{farr2007}) with 90 \si{\metre} resolution (SRTM90) is employed in the retrieval of S2 data (see Sec. \ref{sec:s2}). 

\subsubsection{Land Cover}
\label{sec:landcov}
CORINE Land Cover database\footnote{\url{https://www.data.gouv.fr/fr/datasets/corine-land-cover-occupation-des-sols-en-france/}} was used to filter the land cover types of the region. Only land cover types 321 (grassland), 322 (moorland), 332 (bare rocks), 333 (sparse vegetation) and 335 (glaciers and perennial snow) were considered valid, hence excluding forests, urbanized area, and water bodies from this study since both modelling and satellite retrieval are difficult in such areas \citep{gascoin2019theia}.

\subsection{Snow observations}
\label{sec:obs}

\subsubsection{MODIS observations}
\label{sec:modis}

\begin{figure}[h!]
	\centering
	\subfloat[]{{
		\centering
		\includegraphics[width=70mm]{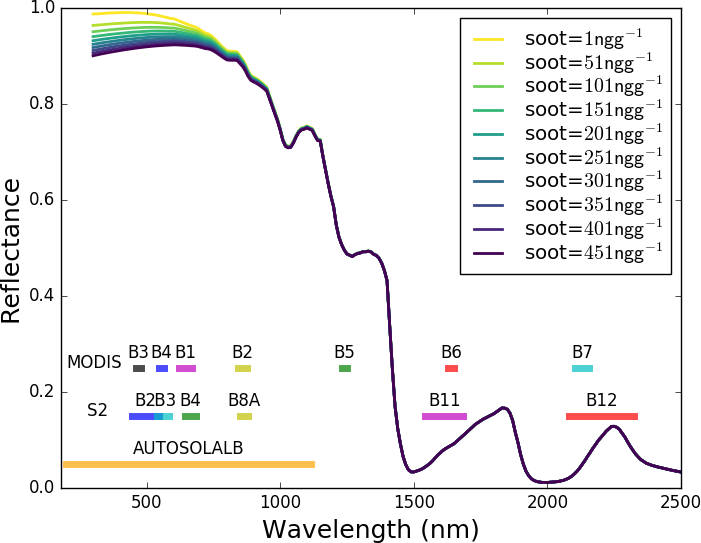}
		\label{fig:tartes_imp}
		}}		
	\subfloat[]{{
		\centering
		\includegraphics[width=70mm]{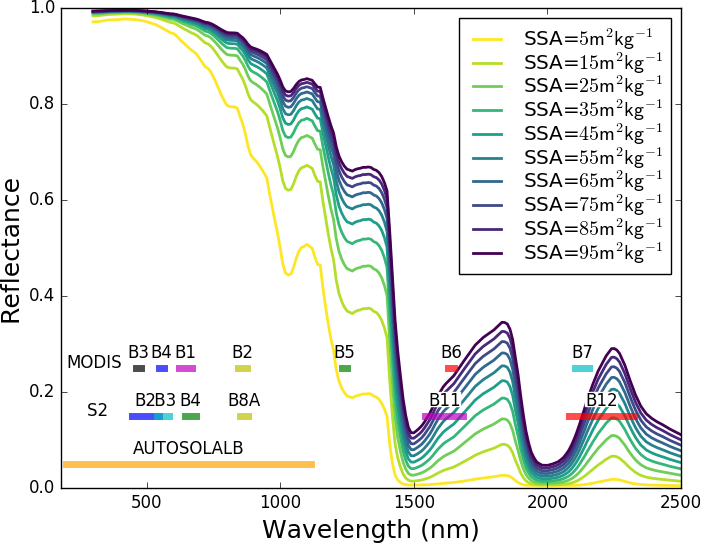}
		\label{fig:tartes_ssa}
		}}	
		
\caption{Computation of snow diffuse reflectances using TARTES for varying soot concentrations (SSA=40\si{\metre^2.\kilogram^{-1}}) (\ref{fig:tartes_imp}) and varying SSA (\ref{fig:tartes_ssa}), for 1 m of 300\si{\kilogram.\meter^{-3}} density uniform snowpack, together with MODIS and S2 spectral bands.\\Source : http://snowtartes.pythonanywhere.com}
	\label{fig:tartes}
\end{figure}

\begin{figure}
	\centering
	\includegraphics[width = 90mm]{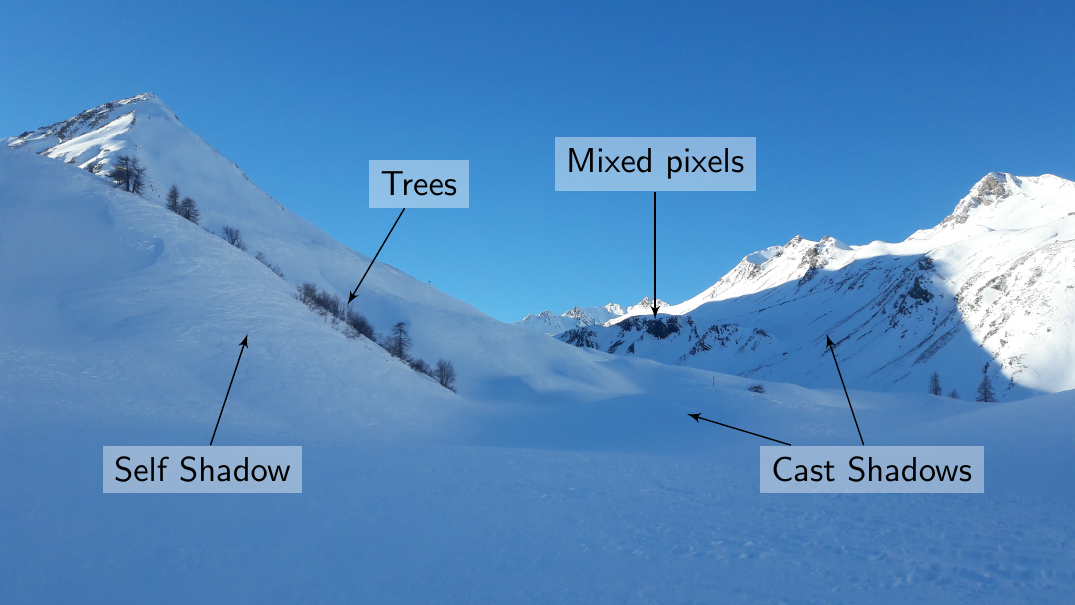}
	\caption{Example of the complexity of the retrieval of reflectance affected by shadows, trees, and mixed snow covers in a complex terrain. (Bertrand Cluzet, Col du Lautaret, December 20\textsuperscript{th} 2017) }
	\label{fig:complex}
\end{figure}

MODIS top of atmosphere radiance in the first seven spectral bands are available at 250 to 500 m spatial resolution depending on the channel (see. Tab. \ref{tab:bands}). As depicted in Fig. \ref{fig:tartes} and Tab. \ref{tab:bands}, reflectance in visible bands (1,3,4) is mostly affected by the impurities content in snow (BC and dust) whereas it depends mostly on SSA in the near-infrared spectral bands (2,5,6,7) \citep{dozier2009interpretation}.\\
We extracted and post-processed these data in a 36x41 km\textsuperscript{2} region (23616 pixels of 250 m resolution, see Fig. \ref{fig:map1}) including the Grandes-Rousses and Col du Lautaret field site during 2013-2014 and 2016-2017 snow seasons with MODImLab retrieval algorithm. In such context of complex terrain, MODImLab retrieval algorithm \citep{sirguey2009b} was shown to outperform other products in many studies \citep{dumont2012b,charrois2013}). Indeed, MODImLab accounts for atmospherical radiative transfer, direct and diffuse contribution, multiple topographical reflection, terrain shading and snow reflectance anisotropy (see. Fig. \ref{fig:complex}).

For mixed pixels, MODImLab's spectral unmixing algorithm computes the reflectance of the snow fraction of the pixel together with a Snow Cover Fraction (SCF). For all the pixels, resulting product is the bi-hemispherical reflectance (accounting in particular  for snow Bidirectional Refletance Density Function (BRDF), \citep{dumont2011}), with 250 m resolution in all bands. MODImLab provides additional masks for shadows (self and cast, see Fig. \ref{fig:complex}) and clouds. For both snow seasons, dates with good geometrical acquisition properties (Sensor Zenithal Angle (SZA) $\leq$ 30\textsuperscript{o}), and clear sky were selected (see Tab. \ref{tab:dates} in Appendix) in order to ensure a maximal accuracy, following \cite{sirguey2016reconstructing} and \cite{charrois2016}.

\subsubsection{Sentinel-2 observations}
\label{sec:s2}
S2 is an ESA-Copernicus satellite program operational since 2016, carrying a multi-spectral high resolution (10-20 m) VIS/NIR sensor with several bands coinciding with MODIS wavelengths (see Tab. \ref{tab:bands} and Fig. \ref{fig:tartes}). Sentinel-2 ground flat bi-hemispherical reflectance products (product FRE, assuming a Lambertian surface) are retrieved by the MAJA processor \citep{hagolle2017maja}, which is similar to MODImLab. Snow masks are retrieved by Let It Snow algorithm\footnote{\url{http://tully.ups-tlse.fr/grizonnet/let-it-snow/blob/master/doc/tex/ATBD_CES-Neige.pdf}} and distributed by Theia Land data center\footnote{CNES.; Gascoin, S.; Grizonnet, M.; Hagolle, O.; Salgues, G. Theia Snow collection, 2017} \citep{gascoin2019theia}. Acquisition is done close to nadir, with $SZA \leq 10\textsuperscript{o}$. Seven clear sky dates were selected during the 2016-2017 snow season (see Tab. \ref{tab:dates} in Appendix).

\subsubsection{In-situ observations}
\label{sec:in-situ}
Autosolalb is a high accuracy instrument measuring snow bi-hemispherical reflectance in the VIS/NIR spectrum (200-1100 nm, 3 nm resolution) including MODIS bands 1-4 \citep{dumont2017situ}. In-situ Autosolalb observations of snowpack bi-hemispherical reflectance were acquired at Col du Lautaret field site (see Fig. \ref{fig:map1} for location) during 2016-2017 winter. The acquisition time step is 12 minutes and acquisition for 2016-2017 winter started on 2017, February 16\textsuperscript{th}. For a given observation time (see Tab. \ref{tab:dates} in Appendix), observation was computed as the mean of all available measurements within +30/-30 minutes and corrected for local slope effects as in \cite{dumont2017situ}.\\

\subsection{Model}
\label{sec:model}
In S2M (SAFRAN-SURFEX/ISBA/Crocus-MEPRA), the Meteo-France operational modelling system of the snowpack, meteorological forcings from SAFRAN analysis \citep{durand1993} are used as inputs to the coupled multilayer ground/snowpack model SURFEX/ISBA/Crocus \citep{vionnet2012}. Ensemble versions for these two components were used here.
\subsubsection{Ensemble of Meteorological Forcings}
\label{sec:forc}
In SAFRAN, a meteorological guess from the NWP model ARPEGE is adjusted with weather observations within each massif on the semi-distributed geometry. Here, in order to represent the uncertainties of this analysis, an ensemble of 35 meteorological forcings was generated by stochastic perturbations on all the meteorological variables of the reference SAFRAN analysis for the Grandes-Rousses. Following \cite{charrois2016}, the magnitude of perturbations was adjusted by a local assessment of SAFRAN errors. SAFRAN does not provide impurities deposition fluxes. Therefore, LAP wet and dry deposition fluxes for BC and dust were extrapolated from MOCAGE chemistry-transport model \citep{josse2004} at Lautaret field site (see Fig. \ref{fig:map1}). For LAP fluxes, \cite{tuzet2017multilayer} showed that the order of magnitude were badly captured by ALADIN-Climate chemistry-transport model \citep{nabat2015dust}, while the timing of events was well captured. Similar behaviour was found with MOCAGE, with an over estimation of BC fluxes in particular. As a consequence, each of the 4 LAP fluxes variables, for each of the 35 members, was multiplied by a constant random factor along the forcing time period, following a lognormal law ($\mu=0.01, \sigma=10$) for BC, and ($\mu=1, \sigma=10$) for dust.\\

\subsubsection{Ensemble of snow models}
\label{sec:escroc}
ESCROC \citep{lafaysse2017multiphysical} is the multiphysical ensemble version of SURFEX/ISBA/Crocus handling 7774 different model configurations. For this study, the last developments of the radiative transfer model TARTES and LAP handling in Crocus were mandatory to properly model the snowpack reflectance (T17 option of radiative transfer, \cite{tuzet2017multilayer}), which were not included in \cite{lafaysse2017multiphysical}. An ensemble of 1944 members using T17 option, so-called "E1tartes" was built for this study, including all the physical options described by \cite{lafaysse2017multiphysical} except for options of solar radiation absorption scheme.

\subsubsection{Model chain}
\label{sec:chain}
\begin{figure}[h!]
	\centering
	\includegraphics[width=140mm]{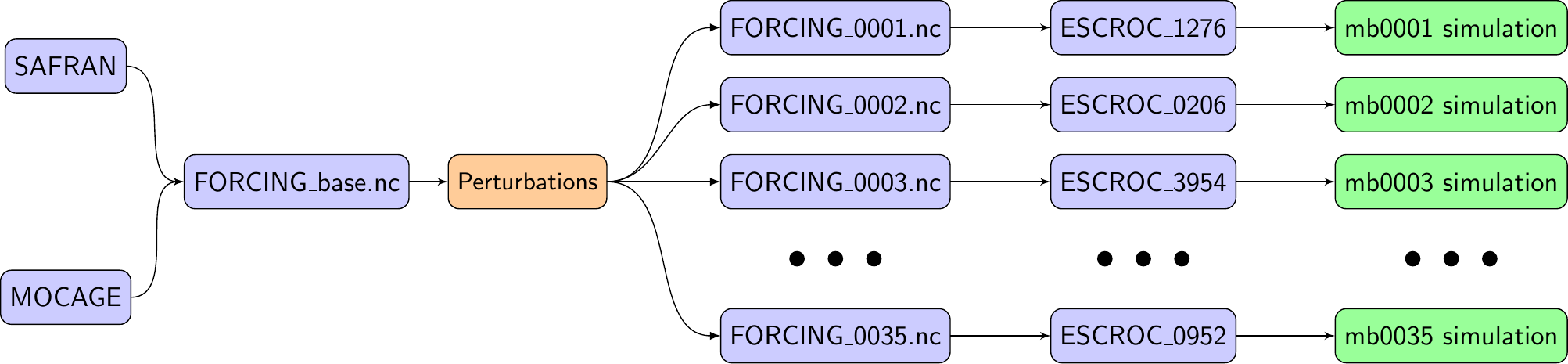}
	\caption{Setup of the ensemble modelling chain.}
	\label{fig:ensemble}
\end{figure}

The ensemble modelling chain setup is summarized in Fig. \ref{fig:ensemble}. At the beginning of a simulation, 35 model configurations are randomly drawn from E1tartes. Each one is associated with a perturbed forcing file to perform the simulation for the whole year, totalling 35 different snowpack simulations. 

\section{Methods}
\label{sec:met}
\subsection{Topographic aggregation}
\label{sec:topoaggr}
An aggregation process is used to adapt the observations to the model semi-distributed geometry with the aim of assimilation. Another added value of the aggregation is to reduce random observation errors and average out features that are not accounted for in the model \citep{hyer2011over}.

\subsubsection{DEM and topographical classification}
\label{sec:demtopo}
In our modelling framework, a topographical class $C_i$ is described by a triplet $(e_i,a_i,s_i)$ where the elevation $e_i \in [600, 900, ..., 3600]$, the aspect $a_i \in [0,45,90,... 315]$ (in degrees, clockwise from North), and the slope $s_i \in [20,40]$ (in degrees). Flat classes are described by a triplet $(e_i, -, 0)$. In our case, there is a total of 187 different topographical classes. For each pixel $p$, a triplet (e,a,s) is computed from the IGN250 DEM (see Sec. \ref{sec:dem}) and thus is attributed to a topographical class. The classification rule is described as follows for tilted classes (Eq. \ref{eq:class}) and for flat classes ( Eq. \ref{eq:classflat}):

\begin{equation}
	p(e,a,s)  \in C_i(e_i,a_i,s_i) \iff \left\{
		\begin{array}{ll}
			e \in [e_i-150, e_i+150[\\
			s \in [s_i-10, s_i+10[\\
			a \in [a_i-22.5, a_i +22.5[\\
		\end{array}
		\right.
	\label{eq:class}
\end{equation}

\begin{equation}
	p(e,a,s) \in C_i(e_i,-,0) \iff \left\{
		\begin{array}{ll}
			e \in [e_i-150, e_i+150[\\
			s < 10\\
		\end{array}
		\right.
	\label{eq:classflat}
\end{equation}

Note that this classification process excludes pixels steeper than 50 degrees were both modelling and remote sensing are unsound. 

\subsubsection{MODIS aggregation} 
\label{sec:aggrmod}
\begin{figure}[h!]
\makebox[\textwidth][c]{
	\centerline{\includegraphics[width = 190mm]{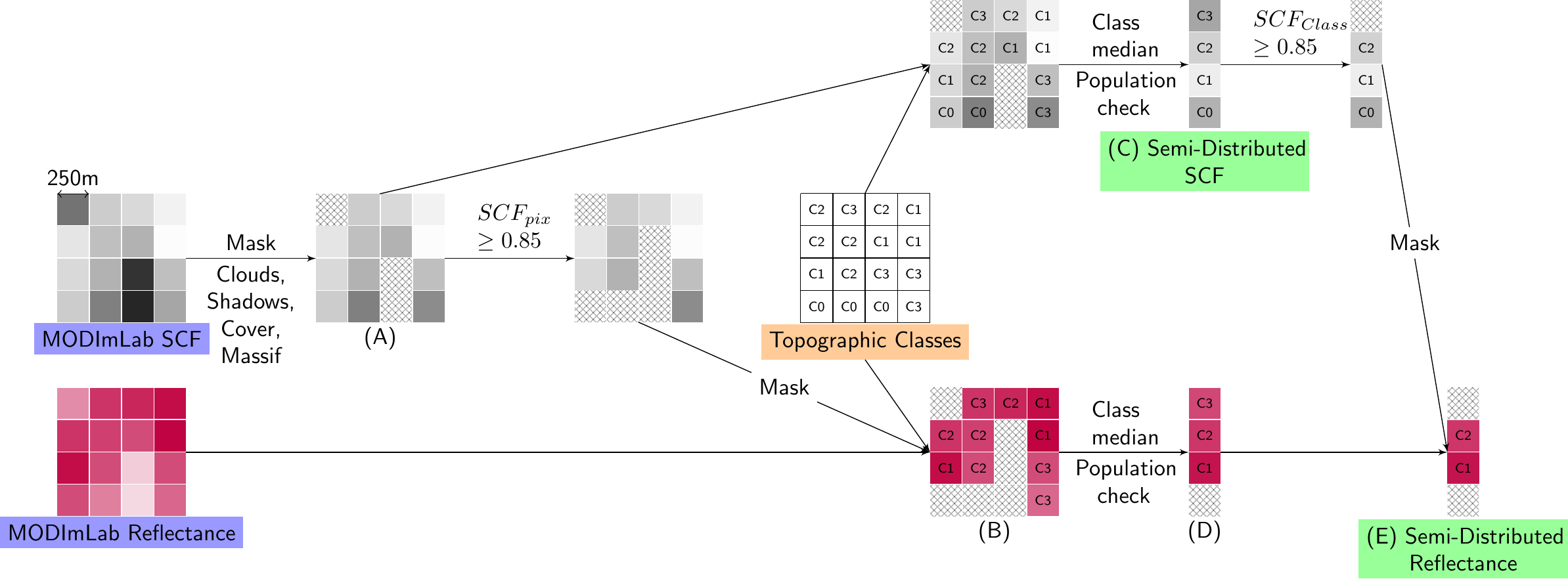}}
	}
	\caption{Flowchart of the conversion of MODImLab products (purple) to semi-distributed data (green), using the Topographical Classification (orange) from Sec. \ref{sec:demtopo}. Masked data are hatched.}
	\label{fig:flow}
\end{figure}

An algorithm is used to aggregate MODIS distributed observations into semi-distributed observations  in order to compare it with model outputs. In this process, a particular attention is paid to the validity and spatial representativeness of the observations, as described in Fig. \ref{fig:flow}. Regarding the validity, pixels with clouds, self/cast shadows, invalid CORINE land covers (see Sec. \ref{sec:landcov}) as well as pixel lying outside the Grandes-Rousses are filtered out (A label in Fig. \ref{fig:flow}). Then for reflectance only, pixels with Snow Cover Fraction $SCF_{pix}$ inferior to 0.85, are discarded (B), since MODImLab reflectance product is less accurate for mixed pixels \citep{mary2013}. The product (B) is referred to as "distributed reflectance".\\
Finally, reflectance and SCF are aggregated into semi-distributed products  by taking the median value within each class. In order to ensure the spatial representativeness of the aggregated observations, classes where the number of valid pixels is below ten and having less than 10\% of pixels with reflectance observations are filtered out in this process (C and D). For the same reason, classes where the average Snow Cover Fraction $SCF_{class}$ is inferior to 0.85 are masked for reflectance in a final step (E).\\

\subsubsection{Sentinel-2 aggregation} 
\label{sec:aggrs2}
\begin{figure}[h!]
\makebox[\textwidth][c]{
	\centerline{\includegraphics[width = 190mm]{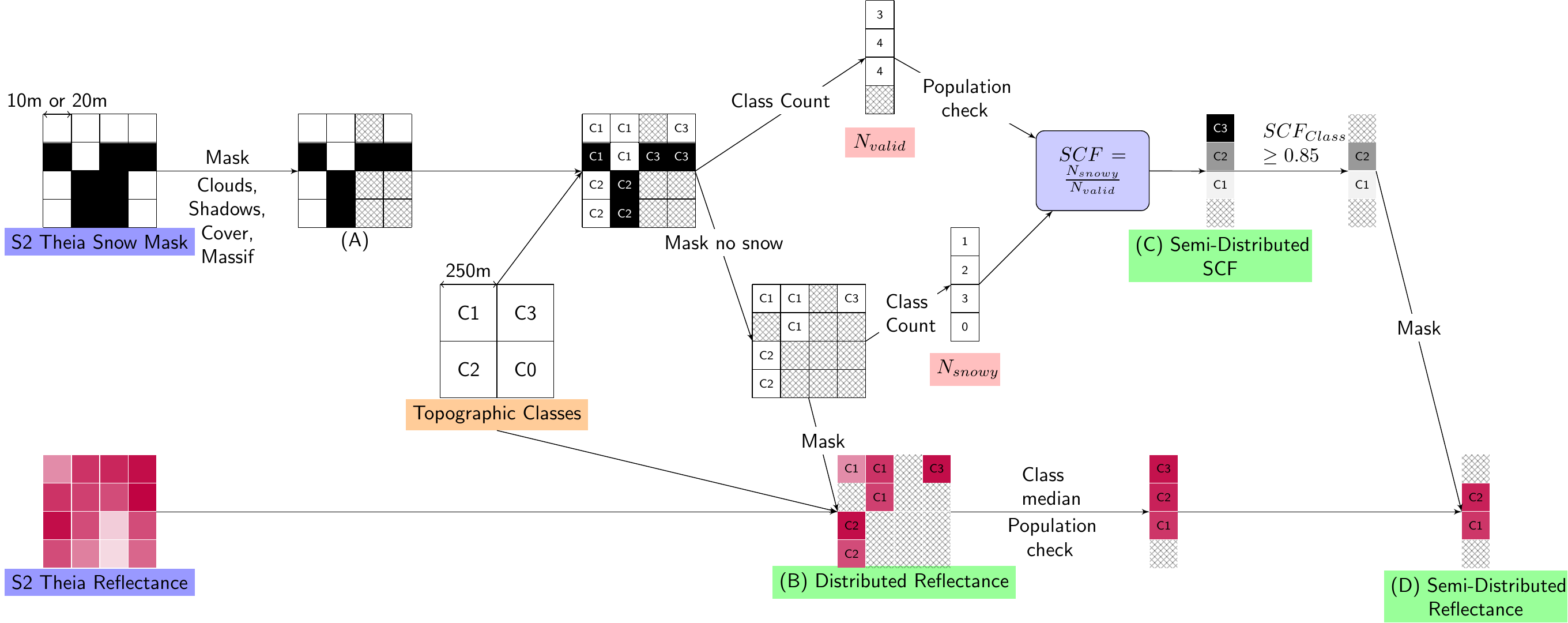}}
	}
	\caption{Flowchart of the conversion of Sentinel-2 products (purple) to semi-distributed data (green), using the Topographical Classification (orange) from Sec. \ref{sec:demtopo}.}
	\label{fig:flow_s2}
\end{figure}

S2 images were aggregated to the semi-distributed geometry in a similar process as for MODIS (see Sec. \ref{sec:aggrmod}), as described in Fig. \ref{fig:flow_s2}. In a first step, a validity masking is performed on Theia L2B Snow Mask using Theia L2A Clouds and Geophysical masks (A). Then, we produce the distributed S2 product (B) by classifying using the IGN250 DEM and discarding non-snow pixels. The aggregated SCF value (D) was here computed as the ratio between snowy and valid populations, when the valid population was above 10 pixels and 10 $\%$ of the total population (as described in the previous paragraph). Finally, aggregated SCF was used to filter the semi-distributed reflectance (D) as in Sec. \ref{sec:aggrmod}.

\subsection{Assessing the feasibility of data assimilation}
\label{sec:feas}
Data assimilation algorithms  generally require that systematical bias between the ensemble and the observations is negligible for a proper functioning \citep{dee1998data}. In addition, for ensemble data assimilation such as the PF, the observation should usually lie within the ensemble envelope, otherwise the algorithm is likely to collapse \citep{charrois2016}. Rank diagrams are commonly used in the ensemble forecasting community to check for both issues by computing the histogram of the position of the observation within the ensemble for all available dates and places \citep{hamill2001}. Furthermore, apart from these considerations, correlations between ensemble and observations timeseries can help quantify the information content from observation and its potential for assimilation \citep{reichle2004global}. If timeseries are weakly correlated, this means that it is likely that observations carry substantial information valuable for the ensemble, but that data assimilation of such different datasets will be a difficult task.\\

In order to assess the potential of applying assimilation algorithms to our spatialized ensemble simulation, a thorough comparison of observed and openloop (i.e. whithout assimilation) simulated reflectances is carried out here : (1) We assess the consistency of the spatial and temporal variations of the ensemble and observations based on two examples (one date and one topographic class). (2) We evaluate the products against in-situ observations, in order to detect systematic biases and errors. (3) We compute Pearson correlations (R) between the ensemble median and semi-distributed observations timeseries in a wide range of topographic classes, to have additional information on the potential of information. (4) We generalize the results by computing rank diagrams, looking for bias and observation position within the ensemble at the same time and over numerous topographic classes and dates.\\


\section{Results}
\label{sec:results}
\subsection{Comparison of observed and simulated variables}
\label{sec:res_compar_refl}

\subsubsection{Spatial comparison on a specific date}
\label{sec:vis_compar}
Fig. \ref{fig:vis_compar} shows maps of NIR semi-distributed reflectance (MODIS band 2) for the two satellite products (MODIS and S2) and the ensemble mean on February 18-19\textsuperscript{th}, 2017. All pixels within the same topographical class are attributed the same value, and in many classes, observations and model are masked out because of shadows.\\
MODIS and S2 remarkably agree on the snowpack extent, while the ensemble mean seems to overestimate it. Both satellite products show on average more contrasted and lower reflectance values than the model. However, MODIS and the model agree on the reflectance dependence on aspect (lower in South-Eastern slopes), contrary to S2.\\

\subsubsection{Ensemble and satellite reflectance timeseries}
\label{sec:res_tc_b}
\begin{figure}[h!]
\makebox[\textwidth][c]{
	\includegraphics[width=190mm]{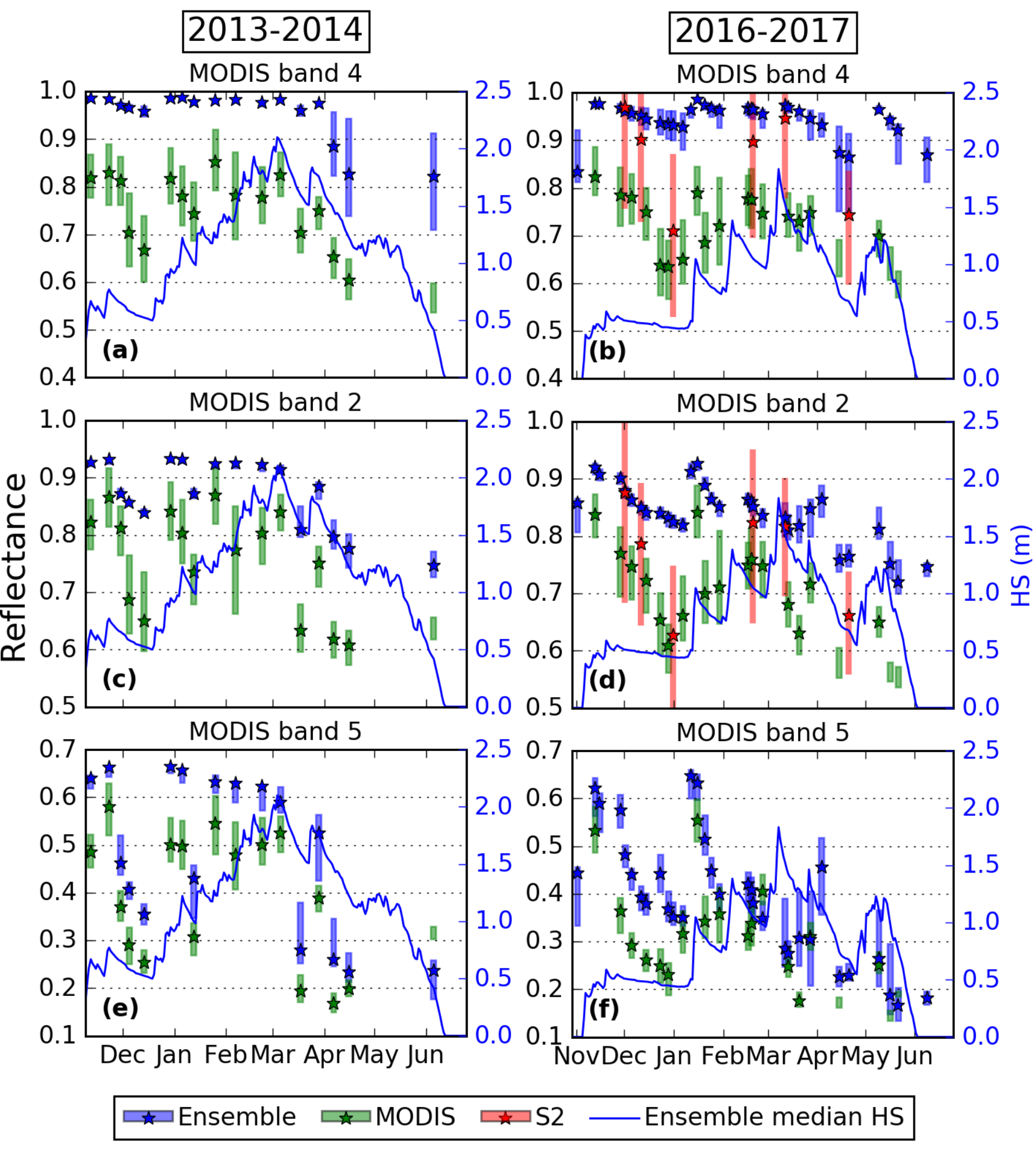}
	}
	\caption{2013-2014 (a,c,e) and 2016-2017 (b,d,f) timeseries of reflectance in MODIS band 4 (a,b), 2 (c,d) and 5 (e,f) for the three different products (ensemble in blue, MODIS in green, S2 in red). The stars denote the median of the ensemble and the semi-distributed satellite products. The boxes shows the ensemble and distributed satellite products quartiles. See Tab.\ref{tab:bands} for the wavelengths and S2 corresponding bands. The blue line denotes the ensemble median Height of Snow (HS).}
	\label{fig:tc_2400_flat}

\end{figure}

Fig. \ref{fig:tc_2400_flat} shows the timeseries of ensemble and observations in MODIS bands 4 (VIS) and 2 and 5 (NIR) for the two snow seasons, in 2400 m flat class. This specific class was chosen here because it is flat, above the tree line and with a long snow covered season, thus easing the comparison all along the snow season. Although there is a strong departure among observations and simulations  (0.1-0.2 in bands 4 and 2, 0.1 in band 5), consistent time variations can be seen between semi-distributed observations (green stars) and the ensemble median (blue stars), for example in December and January of both snow seasons for band 5. For 2013-2014 winter (Fig. \ref{fig:tc_2400_flat} a,c,e), high values of reflectance in all bands during the mid-winter are consistent with the recent snowfall at observation dates during this period (fresh snow has a high SSA, thus a high reflectance as shown in Fig. \ref{fig:tartes}. Decrease in reflectance in all bands from November 2013 to mid December and on January 12\textsuperscript{th} is related with extended periods without snowfall as seen on the HS curve. At the end of the snow season, the snow melt causes a decrease in SSA (i.e. low reflectance in band 2 and 5) due to wet metamorphism \citep{carmagnola2014} . Meanwhile, two dust deposition events (end of February 2014, end of March 2014 in MOCAGE model) can explain drops in band 4 reflectance through an increase in the snowpack surface LAP content. All those events appear in both ensemble and observation timeseries as well as in simulated surface impurities concentrations (not shown). Season 2016-2017 (Fig. \ref{fig:tc_2400_flat}b,d,f) had few, intense snowfall and extended dry periods with clear sky, allowing observe more pronounced reflectance variations.\\
Regarding the ensemble behaviour, in the visible bands, the ensemble Inter-Quartile Range (IQR) (blue boxes) seems generally lower during 2013-2014 winter than in 2016-2017. For all bands, the IQR is reduced after a snowfall (0.01-0.02 in bands 4 and 2, 0.02-0.03 in band 5), and increases with the time elapsed since the last snowfall and all along the melting season (up to 0.1 in bands 4 and 2 and 0.05 in band 5).\\
However, the main feature here is the strong departure between the ensemble and MODIS observations. For almost all dates of both winters, the semi-distributed observation is under all the members of the ensemble in bands  4 and 2. This deviation is smaller in band 5. Note also that the distributed observations IQR (green boxes) is considerable, and notably lower in band 5 (0.02-0.05) than in bands 2 and 4 (0.05-0.1). Regarding S2 observations, (Fig. \ref{fig:tc_2400_flat}b,d), agreement of semi-distributed observations (red stars) with the ensemble is good for fresh snow (2016, December 1\textsuperscript{st}) but a strong departure (0.1-0.2) appears after extended periods without snowfall (2016, December 31\textsuperscript{th} for exambple). Furthermore, the IQR of S2 distributed observations (red boxes) is 2-3 times larger than for MODIS.\\

\subsubsection{Comparison with in-situ measurements}
\label{sec:res_insitu_b}
\begin{figure}[h!]
    \centering
	\includegraphics[width = 90mm]{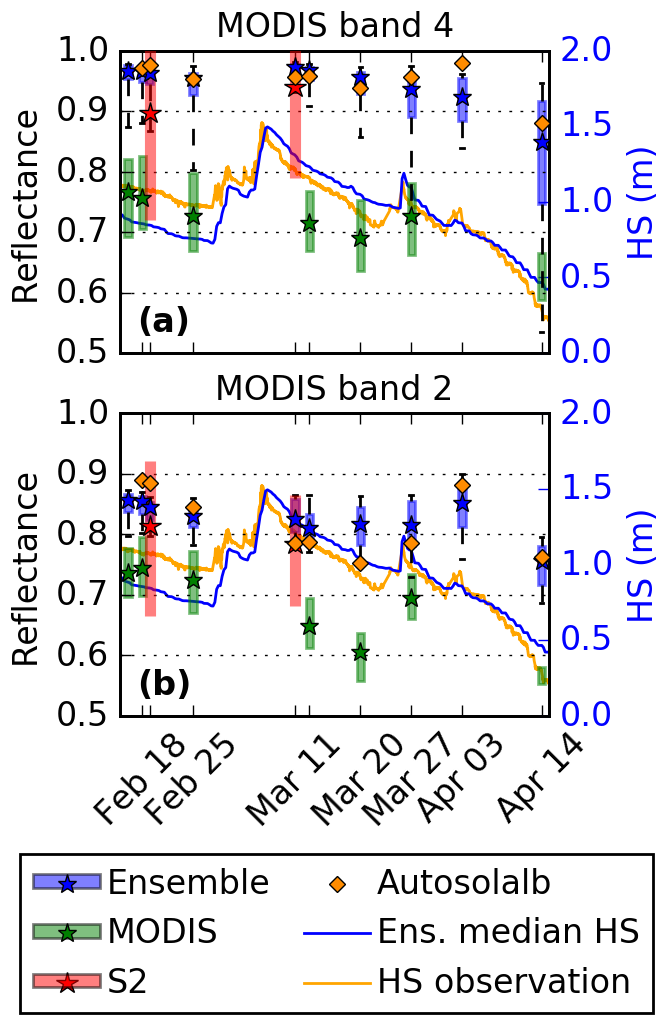}

	\caption{Same as Fig. \ref{fig:tc_2400_flat}, in 2100 m.a.s.l flat class for 2016-2017 winter in MODIS band 4 (a) and 2 (b). In addition, Lautaret data from  Autosolalb (orange diamonds), and observed HS (orange line) are displayed. Note that bars denote the ensemble 5-95\textsuperscript{th}percentiles.}
	\label{fig:tc_solalb}
\end{figure}

%
Comparison with field measurements at Col du Lautaret (Height of Snow (HS) and reflectance in bands 4 and 2) is possible for the 2100 m a.s.l flat class during 2016-2017 winter (see Fig.  \ref{fig:tc_solalb}). First and foremost, there is a strong bias of MODIS observations with respect to in-situ Autosolalb observations (about 0.2 in band 4 and 0.1-0.15 in band 2). However, their time variations reproduce the temporal pattern obtained from in-situ observations for example between March 20\textsuperscript{th} and 27\textsuperscript{th} when an increase of reflectance is occurring in both products.\\
Meanwhile, the ensemble reflectance generally has the same magnitude as the in-situ observations in both bands. In band 4, the in-situ observations lie within the ensemble for fresh snow, for example on February 18\textsuperscript{th}, March 27\textsuperscript{th} and April 3\textsuperscript{rd}. In band 2, reflectance is underestimated by the ensemble for those dates, except on March 27\textsuperscript{th}. In addition, most of the members are overestimating reflectance in both bands during early melt (11\textsuperscript{th} and 13\textsuperscript{th} of March), while the comparison of the ensemble median and in-situ observed HS (blue and orange lines in Fig. \ref{fig:tc_solalb}) show that melt might be underestimated in the model. On March 20th, ensemble band 4 reflectance generally decreases while band 2 increases, together with a light snowfall in the model. Meanwhile, in-situ observations of HS  show that there was no snowfall for this date.\\

\subsubsection{Comparison over all reliable topographical classes}
\label{sec:res_all_b}

To investigate the distribution of this bias over time and space, MODIS observed semi-distributed values were plotted against the ensemble median. We restricted this study to topographical classes where the observation process is the most reliable, i.e. with low probability of being mixed/rocky (20\textsuperscript{o} maximal slope) and with large enough pixel populations over the whole snow seasons (1800-3000 m.a.s.l.). In bands 4 and 2, Figs. \ref{fig:regr_b}a and \ref{fig:regr_b}b show a strong deviation from the 1:1 line. Moreover, the value range in band 4 is much lower in the model (about 0.05) than in the observations (about 0.3). In band 5 (Fig. \ref{fig:regr_b}c), observations and model better align with the 1:1 line.\\

\begin{figure}[h!]
\makebox[\textwidth][c]{
	\subfloat[]{{
		\centering
		\includegraphics[width = 70mm]{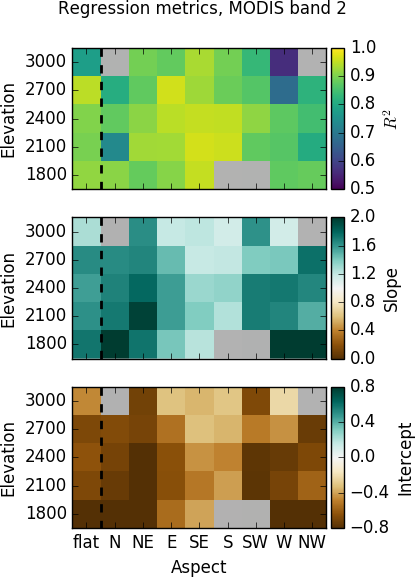}
		\label{fig:statsb2}
		}}
	\subfloat[]{{
		\centering
		\includegraphics[width = 70mm]{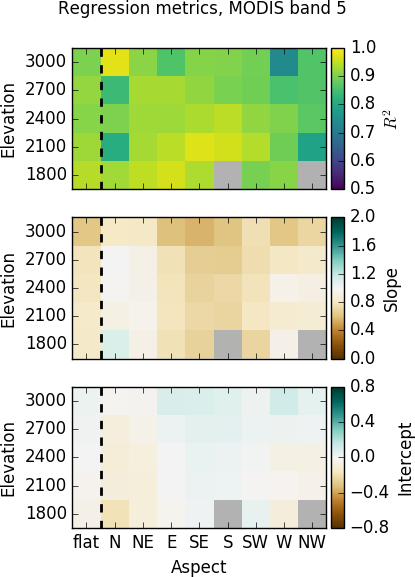}
		\label{fig:statsb5}
		}}
		}
	\caption{Linear regression statistics (upper panel : squared Pearson correlations R\textsuperscript{2}, center panel : regression slope, bottom panel : regression intercept) in band 2 (a) and 5 (b) between the time series of ensemble median and semi-distributed observations for the 45 classes within 1800-3000 m.a.s.l and 0-20 degrees of slope, during 2013-2014  and 2016-2017 snow seasons. Regressions with p-values $>$ 0.01 and less than 6 dates overall are greyed out.}
	\label{fig:statsb}
\end{figure}

In order to refine this analysis over space, linear regressions were systematically carried out between the ensemble median and the semi-distributed observations for each band inside each reliable topographical class (e.g. computing regressions between timeseries of blue stars and green stars in Fig. \ref{fig:tc_2400_flat}). The associated Pearson R\textsuperscript{2}, slope and intercept coefficients are shown in Figs. \ref{fig:statsb2} and \ref{fig:statsb5} for bands 2 and 5. In the absence of model or observational bias, Slope should be close to 1 and Intercept to 0.\\
In band 2, overall high and significant R\textsuperscript{2} (0.75-0.85) are noted. Slope is generally $>$ 1, and Intercept $<$ -0.4. However, regression is close to identity in the sunny slopes (strong dependence on aspect) with higher correlations. Band 5 shows high and significant R\textsuperscript{2} as well (about 0.8-0.9). Slope and Intercept moderately deviates from Identity (Slope $<$ 1).

\subsection{Spectral bands reflectance ratio}
\label{sec:compar_r}
\subsubsection{Timeseries comparison between the model and satellite products}
\label{sec:res_tc_r}
\begin{figure}[h!]
\makebox[\textwidth][c]{
	\includegraphics[width=190mm]{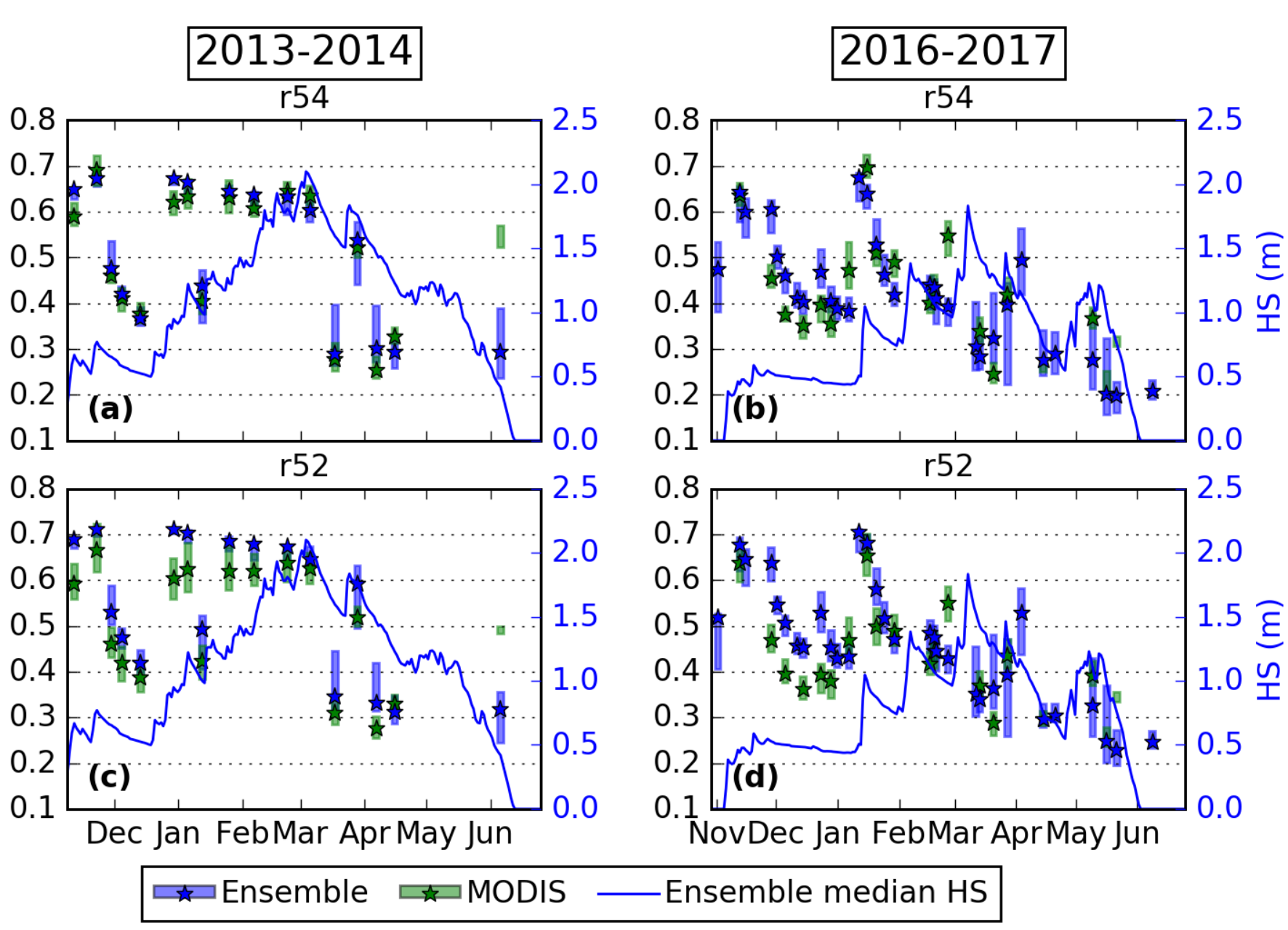}
	}
	\caption{Same as Fig. \ref{fig:tc_2400_flat} for band ratios r54 (a,b) and r52 (c,d).}

	\label{fig:tc_r}
\end{figure}

The bias between observations and model described in Sec. \ref{sec:res_compar_refl} is likely to be problematic for data assimilation. Computing a ratio between the reflectances in two different bands (so-called "band ratio") might reduce this issue.\\
To that aim, the ratios between bands 5 and 4 (r54) and bands 5 and 2 (r52) were computed for MODIS observations. To do so, each ratio was computed on every pixel of the distributed reflectance (label B in Fig. \ref{fig:flow}), and aggregated and masked with the same method as for raw reflectances.\\ 
Fig.\ref{fig:tc_r} shows the temporal evolution of these variables in the 2400 m flat class. Time variations of the ensemble median and semi-distributed observations have compatible values (for example in r54 0.6-0.7 for fresh snow, and 0.25-0.4 in the late season). In about 50\% of the cases, the semi-distributed observation falls within the ensemble IQR (blue boxes) for r54. In addition, note that r52 and r54 signals are very similar, be it in the model or the observations.\\

\subsubsection{Comparison over all the reliable classes}
\label{sec:res_all_r}
Fig. \ref{fig:regr_r} shows the semi-distributed observations against the ensemble medians for the ratios for all the reliable classes and the two snow seasons as in Sec. \ref{sec:res_all_b}. There is no notable systematic bias between the observed ratios and the modelled ones.\\

\begin{figure}[h!]
\makebox[\textwidth][c]{
	\subfloat[]{{
		\centering
		\includegraphics[width = 70mm]{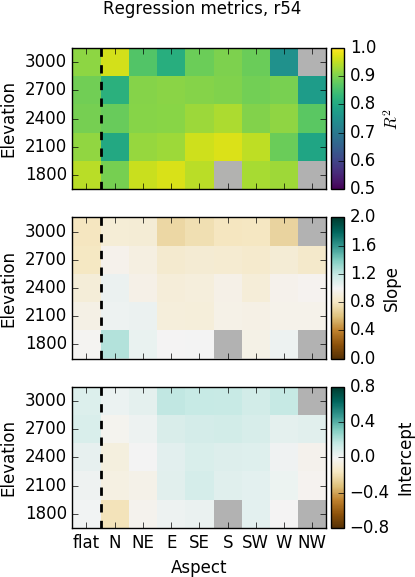}
		\label{fig:stats_r54}
	}}
	\subfloat[]{{
		\centering
		\includegraphics[width = 70mm]{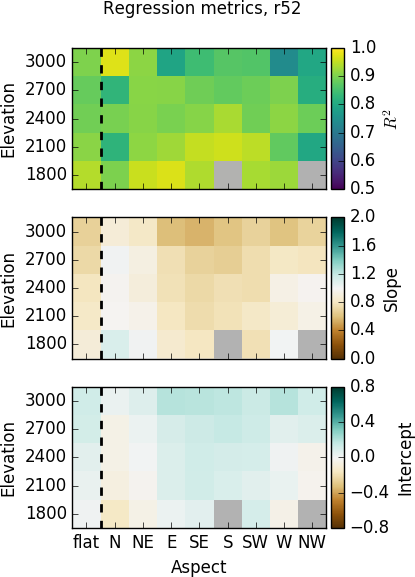}
		\label{fig:stats_r52}
		}}
		}
	\caption{Same as Fig. \ref{fig:statsb2} for r54 (\ref{fig:stats_r54}) and r52 (\ref{fig:stats_r52}).}
	\label{fig:stats_r}
\end{figure}

Statistics of linear regression in Figs. \ref{fig:stats_r54}, and \ref{fig:stats_r52} show high R\textsuperscript{2} values generally above 0.85, similar to those for band 5 in Fig. \ref{fig:statsb5}. More interestingly, regression parameters are now  around identity (Slope=1, Intercept=0) which illustrates the better agreement (no systematic bias) of observations and model for these ratios. While correlation patterns are almost identical for r54 and r52, Slope parameter is generally more departing from identity for r52 than for r54, with a significant dependence on aspect (lower Slopes in sunny aspects).\\

\subsection{Towards assimilation}
\label{sec:disc_towassim}

\begin{figure}[h!]
\centering
    \includegraphics[width = 90mm]{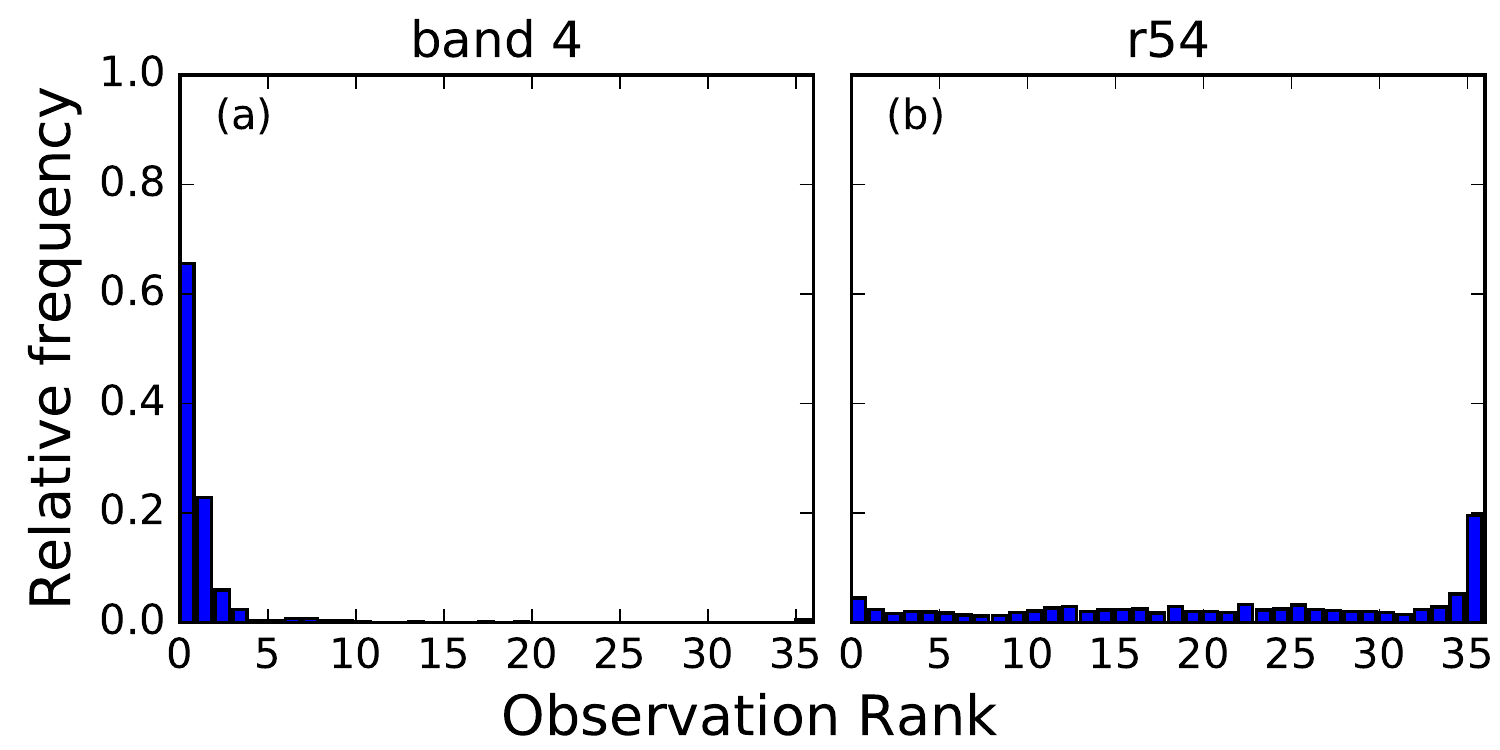}
	\caption{Rank diagrams for the semi-distributed MODIS observations in band 4 (a) and r54 (b) within the ensemble for all classes between 1800 and 3000 m.a.s.l. and between 0 and 20\textsuperscript{o} of slope, and all dates of 2013-2014 and 2016-2017 snow seasons (1009 occurrences).}
	\label{fig:rankdiag}
\end{figure}

Fig. \ref{fig:rankdiag}a shows the rank diagram for the raw reflectance of band 4, over all considered dates and topographical classes of the two snow seasons. In this graph, the observations lie in rank 0 (under all members of the ensemble) about 60 $\%$ of the occurrences, consistently with the negative bias depicted in previous section. 
On the contrary, the rank diagram for band ratio r54 in Fig. \ref{fig:rankdiag}b is highly improved with respect to band 4, the observation being in the ensemble 80 $\%$ of the occurrences. Result is similar for r52 (not shown). Though  overestimation of frequency of ranks 0 (under the ensemble) and 36 (over the ensemble) denote that the ensemble dispersion is insufficient, the rank diagram is flat, all the ranks having similar frequencies.\\

\section{Discussion}
\label{sec:disc}

\subsection{On the relevance of the comparison in the semi-distributed framework}
\label{sec:disc_justif}
The semi-distributed framework was chosen for the comparison between observed and simulated reflectances because it is the basis of the French operational snowpack modelling system, and considering that running this model on a 250m-grid requires about 100 times more computer resources. Since it is quite specific, the different types of errors in observations and simulations in this semi-distributed geometry must be discussed for a correct interpretation of our results. Within a topographical class, observations are affected by (1) natural variability, (2) retrieval errors and (3) classification errors. In particular, DEM errors and resolution have a strong impact in satellite retrievals via shadows and subgrid topography \citep{baba2019effect,davaze2018monitoring}, leading to about $\pm10 \%$ errors in broadband albedo for MODIS data \citep{dumont2012b}. Moreover, S2 data are particularly affected by the three sources, since the retrieval DEM (SRTM90) in the MAJA processor is too coarse to capture the topographic variability at the scale of the pixels (10-20 m) and because the classification is done to an even much coarser scale (IGN250). The resulting intraclass variability of S2 and MODIS is particularly visible in Figs \ref{fig:vis_compar}e, \ref{fig:tc_2400_flat} and \ref{fig:tc_solalb}.\\
However, the resulting distributions of the observations within the classes are reasonably gaussian (see Fig. \ref{fig:histsclass5}), meaning that semi-distributed observations, aggregated by taking the median, should remove random unbiased noises and outliers.\\
From the model point of view, the ensemble approach in this study is expected to satisfactorily assess snowpack modelling errors by the combination of meteorological and multiphysical model ensembles. However the semi-distributed simulations can have a limited spatial representativeness due to the snowpack natural variability, for example when the snow line or rain-snow line lies within the topographic class. In the general case, though, we expect this issue to be of limited importance, in the line with other studies \citep{mary2013}.\\

\subsection{Assets and limits of the satellite products}
\label{sec:disc_asslim}

Since we consider that the observation process is not reliable in shadowed area, we filter out many observations, thus reducing the amount of spatial information available for assimilation. This means that from November to February, North facing slopes will likely not be observed. Therefore, ensemble simulations would not be corrected there during this period, if the assimilation were to be carried out on each topographic class independently. This stresses the need for a spatially coherent data assimilation algorithm, e.g. assimilating all observed topographic classes at the same time, in order to spatially propagate the effect of assimilation and to avoid inconsistent spatial patterns. Furthermore, a spatially comprehensive assimilation of SCF would be needed beforehand to detect topographic classes where the ensemble and observations disagree on the presence of snow and assess where reflectance can be compared, similarly as in \cite{baba2018assimilation}.\\

\begin{figure}[h!]
\centering
    \subfloat{
		\centering	    
	    \includegraphics[width=140mm]{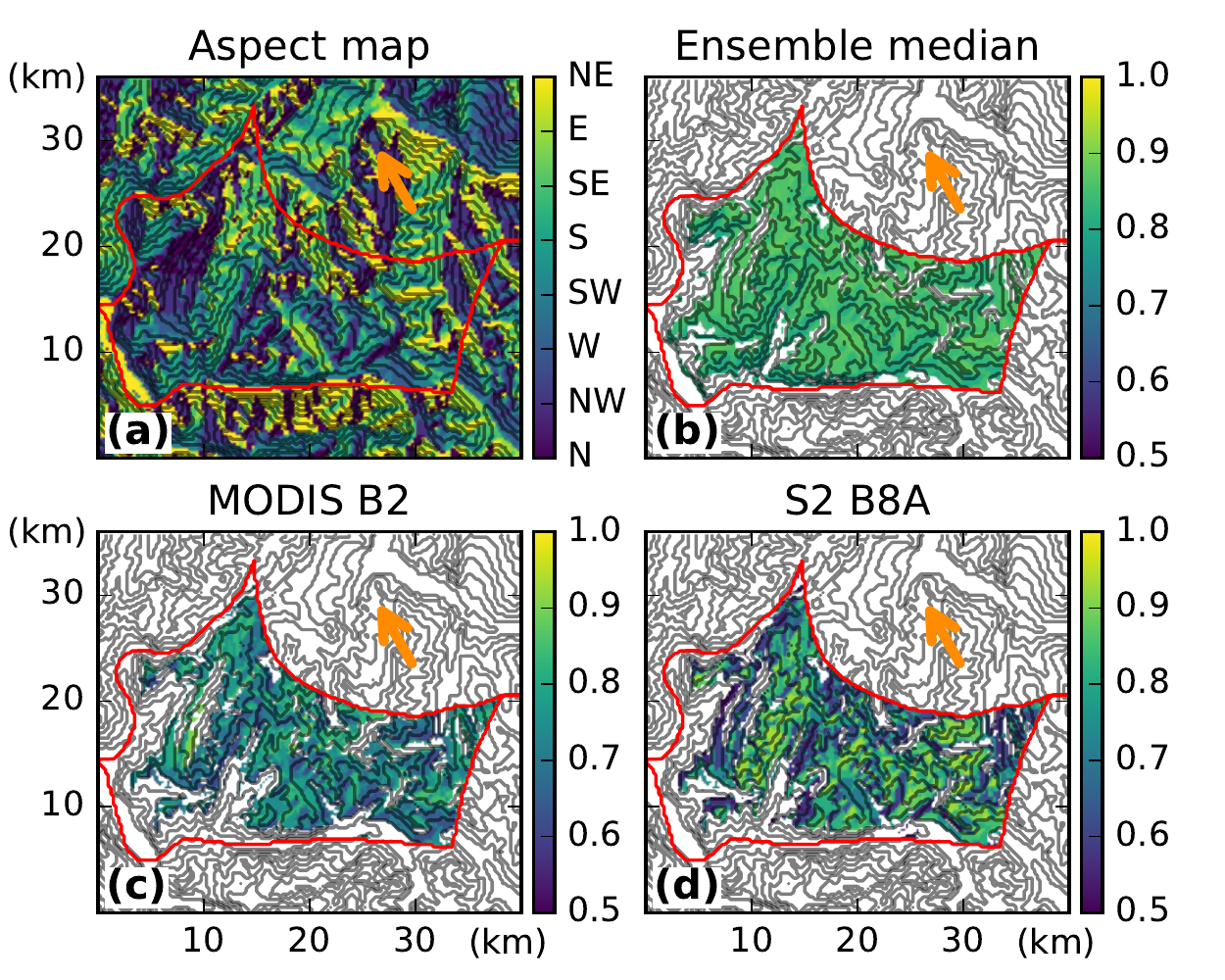}    
	    }\\
	\subfloat{

	    \includegraphics[width=140mm]{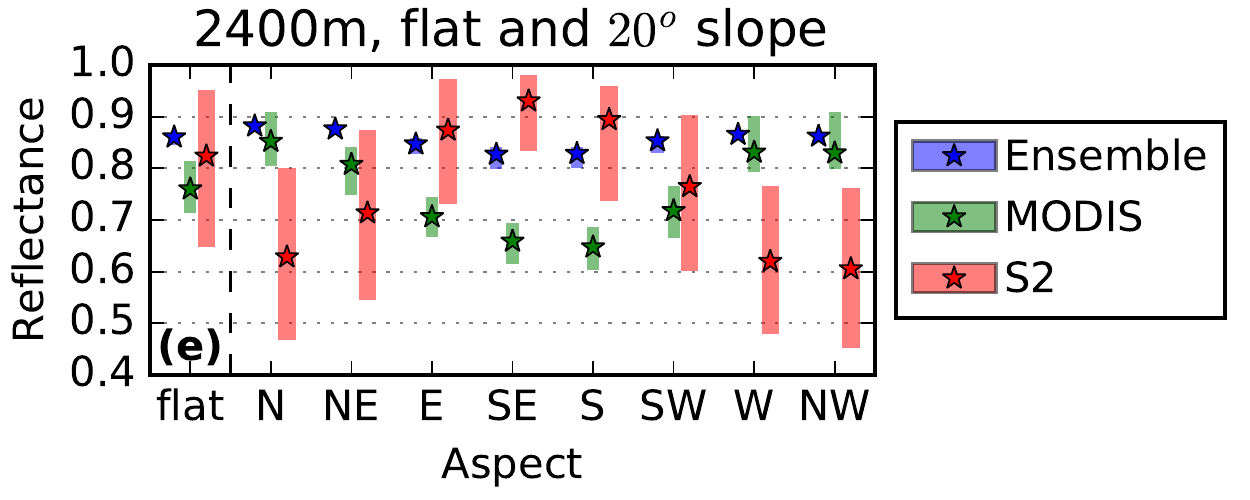}
	}
    \caption{Map of aspect in the Grandes-Rousses (a), and comparison of the 3 reflectance products in the NIR (860nm) on 2017-02-18, 10:00 am: ensemble median (b), semi-distributed MODIS band 2 (c) and S2 Band 8A (2017-02-19, 11:00am) (e). Boxplots (quartiles and medians) for the ensemble (blue), distributed MODIS (green) and S2 (red) in the 2400m, flat and 20\textsuperscript{o} slope classes. On the maps (a-d), the contours denote the model's 300m elevation bands, orange arrows show the approximate sun direction and shadows are masked.}
    \label{fig:vis_compar}
\end{figure}

Observations are also affected by significant errors and biases that are problematic for assimilation. S2 reflectance observations suffer from two significant inconsistencies. (1) The dependence of reflectance on aspect is too strong and unexpected. Higher band 2 reflectance are obtained in South-Eastern slopes where SSA should preferentially decrease owing to sun exposure (causing a decrease in reflectance through enhanced metamorphism) and lower SZA (Fig. \ref{fig:vis_compar}) \citep{warren1982}. (2) Reflectance decrease with time in the absence of snowfall in the early 2016-2017 snow season is too pronounced (Fig. \ref{fig:tc_2400_flat}b and d). These two considerations can be explained by retrieval errors in the MAJA algorithm, probably owing to the representation of topography and atmosphere, which was not specifically designed for snow reflectance retrieval in complex terrain \citep{hagolle2017maja}. In addition, the reflectance retrieval is also affected by the use in MAJA retrieval of a coarse DEM (SRTM90) compared to the native resolution of the data (10-20 m). For all those reasons, improvements in the retrieval of S2 absolute reflectance values is necessary before considering their future assimilation.\\
MODIS reflectance observations also have a strong bias with the model. This bias is unambiguously attributed to MODIS according to the comparison with in-situ observations (Fig. \ref{fig:tc_solalb}). It is much higher than the intraclass variability of the observations and the ensemble IQR. In addition, Figs. \ref{fig:regr_b} and \ref{fig:statsb} show that this bias is well described by a linear function of reflectance which is rather invariant in space and well stable in time.\\
However, MODIS semi-distributed product (median) seems consistent, because : (1) we demonstrate that the median of the observations within the topographical classes is a representative value of the distribution in the general case, (2) reflectance dependence on aspect corresponds to the model one (Fig. \ref{fig:vis_compar}) (3) date-to-date time variations notably match those of the ensemble, (4) these variations sometimes better matches in-situ observations than the ensemble, which proves that their information content is good (Fig. \ref{fig:tc_solalb}, in March). All these considerations give us good confidence in the intrinsic quality and information content of MODIS observations, but a solution to this bias is required for assimilation.\\

\subsection{Assimilating band ratios}
\label{sec:disc_assim}
Biases are a common issue of snowpack remote sensing \citep{veyssiere2019evaluation,balsamo2018satellite} and require a proper estimation or correction before assimilation. Many methods exist in the NWP community to correct for the bias or dynamically estimate it in a data assimilation system \citep{draper2015dynamic,auligne2007adaptive}. However, these methods would require either (1) to assume a non-biased model (2) a representative in-situ reflectance dataset to analyse and model the bias before correcting it on-line (3) extensive, representative, and continuous in-situ observations of snowpack variables to constrain satellite reflectance biases (4) additional data from other satellite sources \citep{balsamo2018satellite}. All of those suffer from limitations owing to the specificities of snowpack modelling and monitoring in a complex terrain, respectively : (1) snowpack reflectance modelling probably suffers from some biases \citep{tuzet2017multilayer} (2) absence of any operational network measuring in-situ snowpack reflectance (3) sparse in-situ snowpack measurements in general (4) lack of reliable reflectance retrieval from other satellite sources (as shown here for S2).\\

Therefore, computing reflectance ratios for assimilation could be an appropriate solution in the current state of the art, because it does not require any assumption on the bias attribution (observations and/or model) and nature. Results show that this method outstandingly allows to unbias the observations using r54 and r52 (Figs. \ref{fig:regr_r} and \ref{fig:stats_r}). Furthermore, band ratios are at the core of snowpack surface properties retrieval from satellites \citep{lyapustin2009,negi2011retrieval,dumont2014contribution,kokhanovsky2018reflectance}. It is not clear, however, whether all the precious information content of reflectance variables is preserved when computing band ratios. Firstly, the correlation of the two unbiased ratios is very high ($\geq 0.9$), as already noted by \citep{lyapustin2009}, and these variables have similar temporal variations than MODIS band 5 (only sensitive to SSA) (see Figs. \ref{fig:tc_2400_flat}e,f and \ref{fig:tc_r}), suggesting that some information on the LAP content might be lost. Since it has been stated that reflectance assimilation requires at least two degrees of freedom, given the dependence of reflectance on LAP and SSA \citep{charrois2016}, further work is required to infer whether these band ratios are varying sufficiently between polluted and pristine snowpacks. Other band combinations, with a higher sensitivity to LAP could also be used (if unbiased), as implemented in \cite{dimauro2015mineral}.\\
Nevertheless, rank diagrams are greatly improved compared to reflectance variables (Fig. \ref{fig:rankdiag}). The obtained almost flat rank diagram for r54 shows that this variable is very likely to fall within the ensemble without any preferential position, for any topographical class and date. This is really encouraging towards spatialized assimilation of such variables.\\

\subsection{Ensemble modelling}
\label{sec:disc_ens}
The remaining underdispersion of the ensemble evidenced by the over representation of the extremal positions in the rank diagrams, could be improved in the near future by a better characterization of the modelling chain uncertainties. (1) Increasing the amplitude of meteorological/impurities fluxes perturbations \citep{charrois2016} or using physical NWP ensemble such as PEARP \citep{descamps2015pearp,vernay2015} could allow to better account for NWP modelling uncertainties and intra-massif variability of weather conditions. (2) Including recent developments in Crocus such as blowing snow within the semi-distributed geometry (SYTRON, \citep{vionnet2018}) (3) Including different impurities scavenging parameter and optical properties configurations within the multiphysical ensemble \citep{tuzet2017multilayer}.\\

Furthermore, adaptations to the presented ensemble modelling chain could make it more suitable for assimilation. First, the ensemble population (N = 35) is small compared to recent local ensemble assimilation attempts in snowpack modelling (e.g. \cite{piazzi2018particle}, \cite{larue2018simulation}, \cite{charrois2016}). However ensemble size must be kept to reasonable values for larger scale operational applications, and scores are not expected to highly depend on ensemble size for openloop simulations \citep{leutbecher2018ensemble}. In addition, though increasing the ensemble population would allow to run several combinations of the forcings with ESCROC members, note that combining each forcing member with only one physical configuration of the model, therefore limiting the combinations, is a current practice in NWP to sample uncertainties \citep{descamps2015pearp}. Secondly, the choice of randomly drawing "N" ESCROC configurations versus carefully building a given subset of "N" members can be discussed. Indeed, \cite{lafaysse2017multiphysical} showed that the ensemble error representativeness could be improved by an appropriate optimized sample of members. However, this sample could not be tested here because it did not include T17 radiative transfer option \citep{tuzet2017multilayer}, mandatory for reflectance modelling. Moreover, site-specific calibrations are expected to be suboptimal when applied over a wide diversity of sites \citep{krinner2018esm}.\\

\section{Conclusions}
\label{sec:concl}
This study investigated the potential for assimilation of MODIS reflectance observations in ensemble snowpack simulations within a semi-distributed framework.\\

First, it is shown that  MODIS observations of reflectance aggregated by topographic classes can be compared with semi-distributed ensemble simulation outputs, and that they convey substantial information content. However, it also clearly appears that MODIS observations are noisy and biased, due to the difficulty of retrieving surface reflectances in a complex terrain. In addition, it seems that S2 reflectance retrieval was affected by even bigger errors.\\

Meanwhile, it seems that the semi-distributed framework is particularly adapted to reflectance assimilation. First, it enables to efficiently remove observational noise thanks to aggregation within topographical classes. It is clear though, that monitoring the substantial intraclass natural variability of reflectance is then out of reach. Furthermore, state-of-the-art distributed snowpack modelling is currently not able to represent this spatial variability either. Reaching this goal would require the use of high resolution meteorological forcings \citep{queno2016snowpack}, and modelling of snow redistribution by wind and gravitation \citep{vionnet2014,mott2010meteorological,freudiger2017snow} in distributed simulations. However, such simulations would require intensive computational resources compared to the semi distributed framework, added to the increase in computational cost due to ensemble forecasting already present here.\\

This study was also the first attempt of spatialized ensemble detailed snowpack modelling using a combination of meteorological and model ensembles. Results showed that the semi-distributed setup is able to represent the associated errors and uncertainties in the modelling of reflectance well, and identified paths to make it more suitable to data assimilation.\\

Therefore, we are confident on the potential for assimilation to take full advantage of reflectance observations and detailed snowpack modelling in such a geometry. However, the remaining strong bias in MODIS semi-distributed reflectance observations prevents from directly assimilating them. A workaround was proposed for MODIS bias by computing ratios of reflectances, a simple method that should preserve the observations information content. We are confident that assimilating such variables is possible and could be beneficial for snowpack modelling in the near future. Furthermore, efforts to improve the retrieval of reflectances in complex terrain must be conducted, in order to reduce retrieval errors and bias, and implement retrieval of other medium-resolution satellite sources such as VIIRS and Sentinel3.\\

\section*{Aknowledgements}
CNRM/CEN is part of Labex OSUG@2020 (investissement d’avenir – ANR10 LABX56). This
study was partly supported by the ANR program ANR-16-CE01-0006 EBONI, LEFE ASSURANCE and APR MIOSOTIS. The authors are grateful to Lautaret staff and Station Alpine Joseph Fourier (SAJF) for ensuring a proper working of the instruments and support for in-situ experiments, P. Sirguey for providing MODImLab code and helpful discussions on retrieval algorithm, and to S. Gascoin, for advice and comments on the handling of Sentinel-2 data. J. Revuelto is supported by a Post-doctoral Fellowship of the AXA research fund (le Post-Doctorant Jesús Revuelto est bénéficiaire d’une bourse postdoctorale du Fonds AXA pour la Recherche Ref: CNRM 3.2.01/17).

\section*{Data and code availability}
The datasets analysed during this study and the code used to produce the figures are available from the corresponding author on request. ESCROC is developed inside the open source SURFEX project (\url{http://www.umr-cnrm.fr/surfex}). While it is not implemented in an official SURFEX release, the code can be downloaded from the specific branch of the git repository maintained by Centre d’Études de la Neige. The full procedure and documentation can be found at \url{https://opensource.umr-cnrm.fr/projects/snowtools_git/wiki/Procedure_for_new_users} and \url{hhttps://opensource.umr-cnrm.fr/projects/snowtools_git/wiki/Data_assimilation_of_snow_observations}. For reproducibility of results, the version used in this work is tagged as cluzetCRST. Processing of the albedo images has been performed using the open-source MODImLab algorithm, (version 1.2.5.d). This algorithm can be accessed by contacting its administrator, P. Sirguey. 

\newpage
\section*{References}
\bibliographystyle{elsarticle-harv} 
\bibliography{biblio/bib_nivo_Bber}

\newpage

\begin{table}[h!]
\centering
\resizebox{\textwidth}{!}{\begin{tabular}{|c|c|c|c|c|c|c|c|}
\hline 
MODIS ID   /S2 ID                           & B3/B2 & B4/B3 & B1/B4 & B2/B8A & B5 & B6/B11 & B7/B12 \\ 
\hline 
Central Wavelength \textit{(nm)} & 469/497 & 555/560 & 645/665 & 858.5/865 & 1240 & 1640/1614 & 2130/2202 \\ 
\hline 
Bandwidth \textit{(nm)}          & 20/100 & 20/45 & 50/40 & 35/33 & 20 & 24/143 & 50/242\\
\hline
Resol. at Nadir \textit{(m)}     & 500/10     & 500/10     & 250/10     & 250/20     & 500     & 500/20 & 500 \\ 
\hline 
Spectral Domain              & VIS     & VIS     & VIS     & VIS/NIR & NIR     & IR  & IR \\ 
\hline
Sensitivity to LAP             & ++      & ++      & ++      & +      &         &    & \\
\hline
Sensitivity to SSA               &    +    &    +    &    +     & ++      & +++      & ++ & ++\\
\hline
Penetration depth \textit{(m)}     &   up to 10-20cm  & a few cm        &  a few cm       & a few cm        &    mm      & mm   & mm\\
\hline 
\end{tabular}}
\caption{MODIS considered spectral band properties together with the closest matching Sentinel-2 band.}
\label{tab:bands}
\end{table}

\newgeometry{left=1cm,bottom=0.1cm, top=1cm, right = 1cm} 
\restoregeometry

\begin{figure}[h!]
\centering
		\includegraphics[width = 140mm]{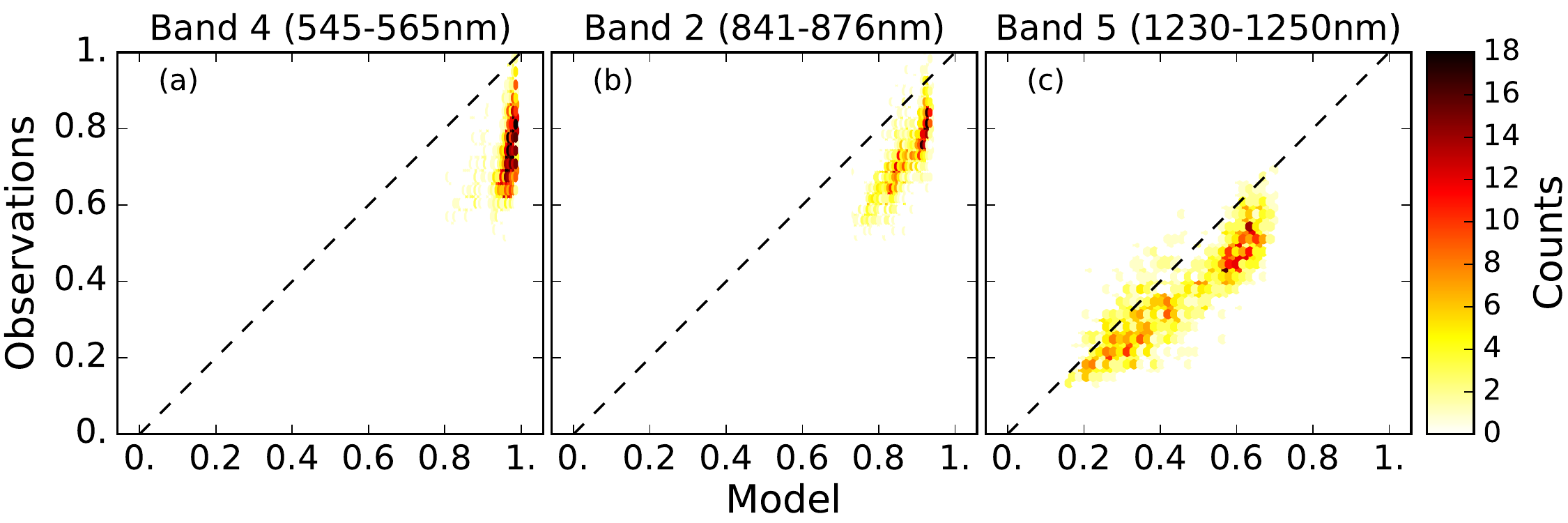}
		\label{fig:regr_b}
\caption{Semi-distributed MODIS observations in band 4 (a), 2 (b) and 5 (c) against ensemble median (density in color), for the 45 topographical classes within 1800-3000m and 0-20 slope, for all the observation dates of 2013-2014 and 2016-2017 snow seasons.}
\label{fig:regr_b}
\end{figure}

\begin{figure}[h!]
\centering
		\includegraphics[width = 90mm]{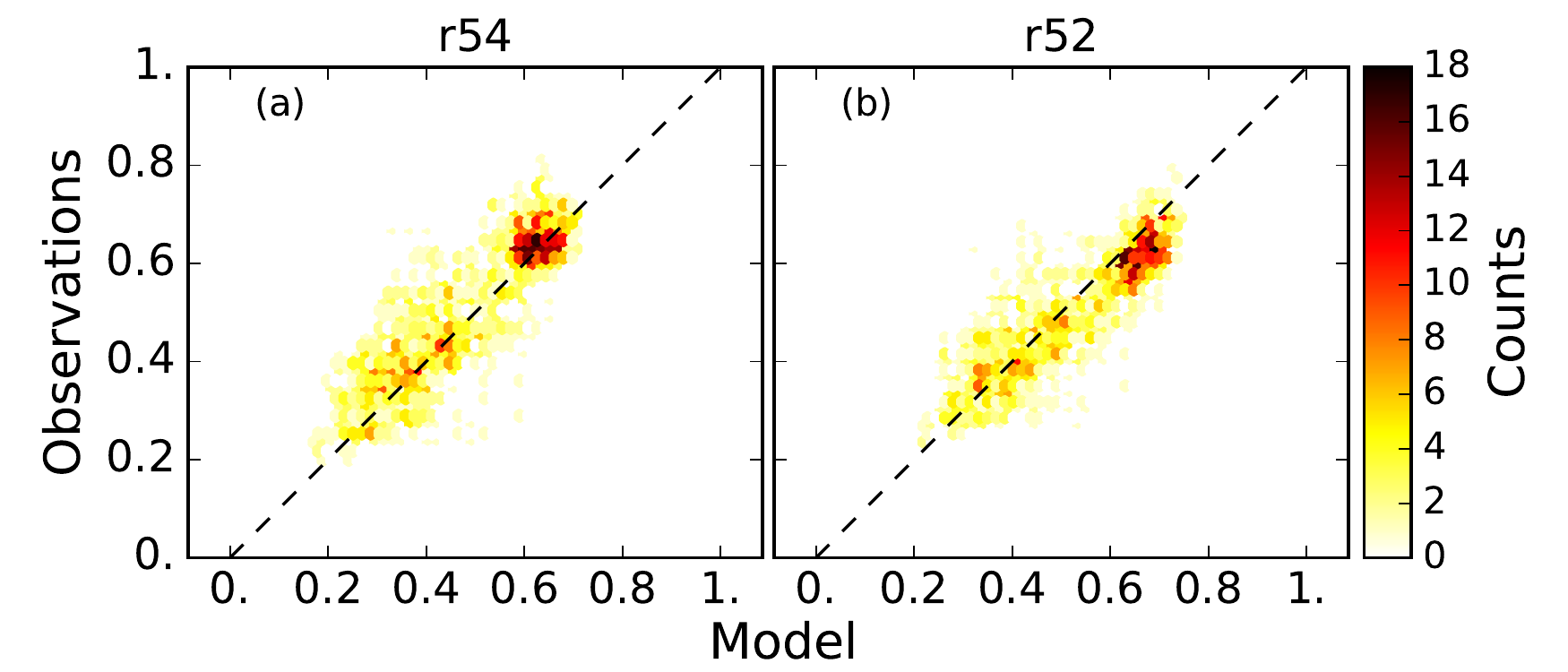}
		\label{fig:regr_r}	
	\caption{Same as Fig. \ref{fig:regr_b}  for r54 (a) and r52 (b).}
	\label{fig:regr_r}
\end{figure}

\clearpage
\appendix
\renewcommand\thefigure{\thesection.\arabic{figure}}
\section{Table of observation dates}
\setcounter{figure}{0}
\setcounter{table}{0}
\label{ap_dates}
\begin{table*}[h!]
\tiny
\centering
\begin{tabular}{|c|c|c|c||c|c|c|c|}                              
\hline                                                           
Date & MODIS & S2 & Autosolalb & Date & MODIS & S2 & Autosolalb\\
\hline                                                           
2013-11-11 11:00 & X &   &   & 2016-12-14 10:00 & X &   &   \\   
2013-11-22 10:00 & X &   &   & 2016-12-23 10:00 & X &   &   \\   
2013-11-29 10:00 & X &   &   & 2016-12-28 11:00 & X &   &   \\   
2013-12-04 11:00 & X &   &   & 2016-12-31 10:00 &   & X &   \\   
2013-12-13 11:00 & X &   &   & 2017-01-06 11:00 & X &   &   \\   
2013-12-29 11:00 & X &   &   & 2017-01-11 11:00 & X &   &   \\   
2014-01-05 11:00 & X &   &   & 2017-01-15 11:00 & X &   &   \\   
2014-01-12 11:00 & X &   &   & 2017-01-20 11:00 & X &   &   \\   
2014-01-25 10:00 & X &   &   & 2017-01-24 10:00 & X &   &   \\   
2014-02-06 11:00 & X &   &   & 2017-01-29 11:00 & X &   &   \\   
2014-02-22 11:00 & X &   &   & 2017-02-16 11:00 & X &   & X \\   
2014-03-05 10:00 & X &   &   & 2017-02-18 10:00 & X &   & X \\   
2014-03-17 11:00 & X &   &   & 2017-02-19 11:00 &   & X & X \\   
2014-03-28 10:00 & X &   &   & 2017-02-25 10:00 & X &   & X \\   
2014-04-06 10:00 & X &   &   & 2017-03-11 11:00 &   & X & X \\   
2014-04-15 10:00 & X &   &   & 2017-03-13 10:00 & X &   & X \\   
2014-06-05 11:00 & X &   &   & 2017-03-20 11:00 & X &   & X \\   
\cellcolor{blue!25}Winter 2016-2017 & \cellcolor{blue!25}  &  \cellcolor{blue!25} & \cellcolor{blue!25}  & 2017-03-27 11:00 & X &   & X \\   
2016-11-01 11:00 &   & X &   & 2017-04-03 11:00 &   & X & X \\   
2016-11-12 10:00 & X &   &   & 2017-04-14 10:00 & X &   & X \\   
2016-11-15 11:00 & X &   &   & 2017-04-20 10:00 & X &   & X \\   
2016-11-28 10:00 & X &   &   & 2017-05-09 10:00 & X &   &   \\   
2016-12-01 11:00 &   & X &   & 2017-05-16 10:00 & X &   &   \\   
2016-12-05 11:00 & X &   &   & 2017-05-21 11:00 & X &   &   \\   
2016-12-11 11:00 &   & X &   & 2017-06-08 11:00 & X &   &   \\   
\hline 
\end{tabular} 
\caption{Summary of observation dates for MODIS, S2 and Autosolalb sensors over 2013-14 and 2016-2017 winters. Time is given for the corresponding closest model output time step (hour).}
\label{tab:dates}
\end{table*}

\section{Intraclass distribution of observations}
\setcounter{figure}{0}
\setcounter{table}{0}
\label{ap_var}

\begin{figure*}[h!]
\centering
\includegraphics[width =140mm]{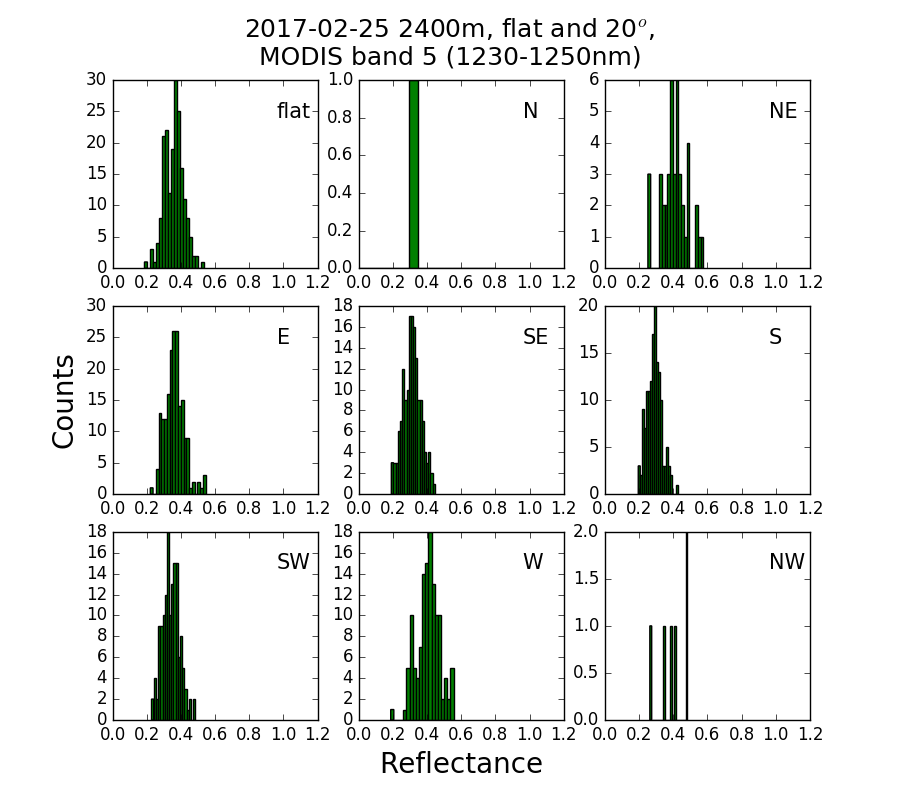}
\caption{Histograms of MODIS band 5 reflectance in flat and 20\textsuperscript{o} slope classes at 2400m on 2017, February the 25\textsuperscript{th}, 10:40am.}.
 \label{fig:histsclass5}
\end{figure*}

\end{document}